\documentclass[preprint]{aastex6}
\usepackage{amsmath}
\usepackage{natbib}
\usepackage{epstopdf}
\usepackage{float}
\usepackage[caption = false]{subfig}
\usepackage{graphicx}
\usepackage{listings}
\usepackage{rotating}

\newcommand{\lapprox} {\, \lower3pt\hbox{$\sim$}\llap{\raise2pt\hbox{$<$}}\,}
\newcommand{\gapprox} {\, \lower3pt\hbox{$\sim$}\llap{\raise2pt\hbox{$>$}}\,}

\begin{document}

\title{Heating and cooling of coronal loops with turbulent suppression of parallel heat conduction}

\author{Nicolas Bian\altaffilmark{1,2}, A. Gordon Emslie\altaffilmark{2}, Duncan Horne\altaffilmark{1}, and Eduard P. Kontar\altaffilmark{1}}

\altaffiltext{1}{School of Physics \& Astronomy, University of Glasgow, Glasgow G12 8QQ, Scotland, UK \\}
\altaffiltext{2}{Department of Physics \& Astronomy, Western Kentucky University, Bowling Green, KY 42101 (nicolas.bian@wku.edu,emslieg@wku.edu)}

\begin{abstract}
Using the ``enthalpy-based thermal evolution of loops'' (EBTEL) model, we investigate the hydrodynamics of the plasma in a flaring coronal loop in which heat conduction is limited by turbulent scattering of the electrons that transport the thermal heat flux. The EBTEL equations are solved analytically in each of the two (conduction-dominated and radiation-dominated) cooling phases. Comparison of the results with typical observed cooling times in solar flares shows that the turbulent mean free-path $\lambda_T$ lies in a range corresponding to a regime in which classical (collision-dominated) conduction plays at most a limited role. We also consider the magnitude and duration of the heat input that is necessary to account for the enhanced values of temperature and density at the beginning of the cooling phase and for the observed cooling times. We find through numerical modeling that in order to produce a peak temperature $\simeq 1.5 \times 10^7$~K and a 200~s cooling time consistent with observations, the flare heating profile must extend over a significant period of time; in particular, its lingering role must be taken into consideration in any description of the cooling phase. Comparison with observationally-inferred values of post-flare loop temperatures, densities, and cooling times thus leads to useful constraints on both the magnitude and duration of the magnetic energy release in the loop, as well as on the value of the turbulent mean free-path $\lambda_T$.
\end{abstract}


\section{Introduction}

Hard X-ray imaging spectroscopy observations from \emph{RHESSI} \citep[e.g.,][]{2004ApJ...603L.117V,2008A&ARv..16..155K,2012ApJ...755...32G} have shown that coronal loop-top hard X-ray sources are common.  Such sources require that the bremsstrahlung-producing electrons are both accelerated in, and subsequently confined to, the corona. \cite{2013A&A...551A.135S} have further shown that, even in events with both coronal and footpoint sources, the number of electrons remaining in the corona relative to those that precipitate downwards to the chromosphere is larger than would be expected on the basis of a purely collisional \citep{1972SoPh...26..441B,1978ApJ...224..241E} model of electron energy transport.  These observations have led various authors \citep[e.g.,][]{2014ApJ...780..176K, 2017ApJ...835..262B} to consider mechanisms that act to more effectively confine accelerated electrons in the corona, in particular the possibility that turbulent fluctuations in the ambient magnetic field act to enhance the angular scattering rate and so suppress the rate of escape of non-thermal electrons from the coronal acceleration region. \cite{2017PhRvL.118o5101K} have recently presented observations of broad soft X-ray spectral lines formed near the top of a flaring loop, showing that a significant level of turbulent bulk motion is indeed present near the electron acceleration region.  They thus infer that turbulence may well be a key element in the transfer of energy from the stressed magnetic field to the accelerated particles.  The presence of such turbulent fluctuations in the ambient medium will also have an impact on the transport properties of the plasma, such as its thermal and electrical conductivities \citep{2016ApJ...824...78B}.

A long-standing problem in solar physics \citep[e.g.,][]{1980sfsl.work..341M} is that observed cooling times of soft-X-ray-emitting post-flare coronal loops are much longer than expected from a model in which cooling proceeds by collision-dominated \citep{1962pfig.book.....S} conduction.  The recent comprehensive study of loop cooling times by \cite{2013ApJ...778...68R} explores this issue in some detail. Figure~1 of that paper shows a typical cooling trend, deduced from the times of peak intensity of spectral lines that are formed at progressively lower temperatures. Column~4 of their Table~1 provides measurements, for 72 events, of the time it takes to cool from a temperature $T_0 \simeq 1.5 \times 10^7$~K (the peak formation temperature of the 192 \AA\ Fe XXIV line observed by the SDO EVE experiment \citep{2012SoPh..275..115W}.
Column~5 of Table~1 in \cite{2013ApJ...778...68R} also gives the estimated cooling time using the static, classical conduction, pure cooling model of \cite{1995ApJ...439.1034C}. Table~1 and Figure~5 of \cite{2013ApJ...778...68R} show that the observed cooling times are systematically higher than those of the \cite{1995ApJ...439.1034C} model, leading them to suggest, consistent with earlier work \citep[e.g.,][]{1980sfsl.work..341M}, that additional heat input is present during the cooling phase.

These results provide a powerful motivation to consider the possible limitation of heat conduction by turbulence as an alternative (or, as we shall discover below, additional) explanation for the long observed cooling times. \cite{2016ApJ...824...78B} obtained analytical expressions for the conductive heat flux (and current density) via a Chapman-Enskog expansion of the electron kinetic equation where diffusion in pitch-angle space includes both collisional and turbulent processes, and proceeded \citep{2016ApJ...833...76B} to consider the impact of this turbulent limitation of the thermal heat flux on the post-impulsive-phase cooling of flare loops.  However, this seminal analysis assumed, for simplicity, a static loop and hence neglected mass motions, which \citep[e.g.,][]{1976SoPh...50..133C} are of considerable importance in determining the overall evolution of the loop plasma.

Modeling the hydrodynamic evolution of the plasma in a post-flare coronal loop requires consideration of a complicated interplay amongst the spatial and temporal distributions of deposited energy, the resultant heating and thermal evolution, and the bulk mass motions driven by the pressure gradients established by the heating. Over the past few decades, a series of numerical models of progressively increasing complexity \citep[e.g.,][]{1983SoPh...86..147P,1984ApJ...279..896N,1986SoPh..103...47M,1989ApJ...341.1067M,2005ApJ...630..573A,2015ApJ...809..104A}, have been developed.  An early ``benchmarking'' study of these various models \citep{1984MmSAI..55..811K,1986epos.conf..7.1K} showed that while there were some subtle variations between the results of various numerical methods when addressing the same problem, the general properties of the solutions, particularly regarding the overall partitioning of energy amongst thermal energy and hydrodynamic motions, tended to be in substantial agreement.  This realization eventually led to the development of ``zero-dimensional'' (0-D) models of flaring loops, in particular the ``enthalpy-based thermal evolution of loops'' (EBTEL) model of \citet{2008ApJ...682.1351K} and \citet{2012ApJ...752..161C} that consists of simple ordinary differential equations that describe the temporal evolution of loop-averaged quantities such as density, temperature, and pressure. Despite their huge simplicity relative to the system of simultaneous nonlinear partial differential equations that describe even one-dimensional hydrodynamic models (a simplicity that results in orders of magnitude less computing time), the EBTEL model nevertheless accurately describes, for a wide variety of scenarios \citep{2012ApJ...758....5C}, the time evolution of the spatially-averaged temperature, pressure, and density within a heated coronal loop.

In this paper we therefore extend the results of \cite{2016ApJ...833...76B} by including, through EBTEL modeling, the effect of mass motions on the post-flare cooling of coronal loops in which turbulent scattering leads to a significant reduction in the redistribution of energy due to thermal conduction. We shall also consider the effects of the heating function that establishes the temperature and density of the loop at the beginning of the cooling phase, and, as we shall discover below, must persist into the cooling phase, consistent with the conclusions of \cite{1980sfsl.work..341M} and \cite{2013ApJ...778...68R}.  Despite some rather obvious typographical errors in Table~1 of \cite{2013ApJ...778...68R} (e.g., loop lengths of order 1~cm, densities of order $10^{12}$~cm$^{-3}$), the order of magnitude of the observed cooling times given in their Table~1, further supported by their Figures~2 and~5, shows that the mean cooling rate for the 72 observed flares is $\simeq 3.5 \times 10^4$~K~s$^{-1}$, with a significant skewness to much longer cooling times (smaller cooling rates).  This corresponds to cooling from $T_0 \simeq 1.5 \times 10^7$~K to $1 \times 10^7$~K in about $150$~s or more, and we therefore choose to constrain our modeling by requiring that the loop temperature peak at $\simeq 1.5 \times 10^7$~K, and that it then cools from that temperature to $1 \times 10^7$~K in $\simeq 200$~s.

In Section~\ref{ebtel-equations} we present the EBTEL model equations, incorporating turbulent suppression of heat conduction. In Section~\ref{cooling-regimes-overview} we find analytic expressions for the temperature and density evolution in both conductive and radiative regimes of cooling. In Section~\ref{cooling-scenarios} we apply these results to the cooling of a post-flare loop, following the evolution of the loop temperature and density from starting values $T_0 = 1.5 \times 10^7$~K and $n_0 = 10^{10}$~cm$^{-3}$, for different values of the parameter $\lambda_T$ that characterizes the mean free path associated with the turbulent scattering.  In Section~\ref{heating-function} we turn our attention to the characteristics, particularly the intensity and duration, of the heat input that drives the loop to enhanced values of temperature and density at the start of the flare.  We find that in order to provide enough pressure and density for the loop to be visible in X-rays and EUV without unreasonably high peak temperatures, the heating profile must extend over a significant period of time {\it and therefore must be taken into consideration in any description of the cooling phase}. Comparison with observationally-inferred values of post-flare loop temperatures, densities, and cooling times thus leads to useful constraints on both the magnitude and duration of the magnetic energy release. It also demands an upward adjustment over previous estimates \citep{2016ApJ...833...76B} of the value of $\lambda_T$, bringing it into closer agreement with the earlier estimates of \cite{2014ApJ...780..176K} based on hard X-ray observations.  We conclude that not only is continuing heat input to the corona necessary during the loop cooling phase, but also that thermal conduction is significantly reduced below its purely collisional value, to the extent that at no time does collision-dominated thermal conduction play a dominant role in the cooling of post-flare coronal plasma.

\section{EBTEL Equations including turbulent suppression of heat conduction}\label{ebtel-equations}

\subsection{Form of the thermal conductive flux}

We consider the flare volume to be filled by a strong magnetic field, so that cross-field energy and momentum transport may be considered negligible. We model the thermal conductive flux (erg~cm$^{-2}$~s$^{-1}$) along the loop ($z$-direction) as

\begin{equation}
F_C = -\kappa \, \frac{dT}{dz} \,\,\, ,
\end{equation}
so that the thermal flux is proportional to the local temperature gradient according to Fourier's law \citep[see, however,][]{2017ApJ...835..262B}. A simple dimensional analysis in this case \citep[see, e.g., ][]{2016ApJ...824...78B}), shows that the thermal conductivity coefficient $\kappa$ (erg~cm$^{-1}$~s$^{-1}$~K$^{-1}$) is proportional to the mean free path $\lambda$ (cm):

\begin{equation}\label{kappa-general}
\kappa = \frac{2n k_B (2k_BT)^{1/2}}{m_e^{1/2}} \, \lambda \,\,\, ,
\end{equation}
where $T$ (K) is the electron temperature, $n$ (cm$^{-3}$) is the ambient density, $m_e$ (g) is the electron mass and $k_B = 1.38 \times 10^{-16}$~erg~K$^{-1}$ is the Boltzmann constant.
The mean free path $\lambda$ is sensitive to the nature of the scattering process. In the case of scattering via Coulomb interactions \citep[see, e.g.,][]{1962pfig.book.....S}, the corresponding (collisional) mean free path $\lambda_C$ for an electron with the thermal speed $v_{te} = \sqrt{2 k_B T/m_e}$ is

\begin{equation}\label{lambdaei}
\lambda_{ei} \equiv \lambda_C(v=v_{te}) =
\frac{(2k_B T)^2}{2\pi e^4 \ln \Lambda \, n} \simeq 10^4 \, \frac{T^2}{n} \,\,\, ,
\end{equation}
where $e$ (esu) is the electronic charge and $\ln\Lambda \approx 20$ is the Coulomb logarithm. This yields the usual expression for Spitzer conductivity:

\begin{equation}\label{eq:ks}
\kappa_C = \frac{k_B \, (2k_BT)^{5/2}}{\pi m_e^{1/2}e^4 \ln\Lambda} \equiv \kappa_{0} \, T^{5/2} \simeq 1.7 \times 10^{-6} \, T^{5/2} \,\,\, ,
\end{equation}
a quantity that is independent of density and quite strongly dependent on temperature.

However, as discussed by \cite{2016ApJ...824...78B}, the parallel transport of energy by thermal conduction may be strongly affected by inhomogeneities in the guiding magnetic field.  Indeed, such inhomogeneities are required for the dissipation of magnetic energy on a time scale much smaller than the collisional time-scale (which, for coronal conditions, is much too long to account for observed impulsive phase rise times during flares). The presence of a spectrum of magnetic field fluctuations within the loop gives rise to an additional source of angular scattering for electrons, hereafter referred to as ``turbulent scattering,'' with an associated (velocity-independent) mean free path

\begin{equation}
\lambda_T = \lambda_B \, \left ( \frac{\delta B_\perp}{B_0} \right )^{-2} \,\,\, ,
\end{equation}
where $\lambda_B$ is the magnetic correlation length and $\delta B_{\perp}$ is the magnitude of the fluctuations perpendicular to the background field $B_{0}$ (Gauss). For such a mean free path, Equation~(\ref{kappa-general}) shows that the thermal conductivity coefficient is given by

\begin{equation}\label{eq:kt}
\kappa_T = \frac{2 n k_B(2k_BT)^{1/2}}{m_e^{1/2}} \, \lambda_{T} \simeq 1.5 \times 10^{-10} \, \lambda_T \, n \, T^{1/2} \,\,\, .
\end{equation}
In contrast with collision-dominated (Spitzer) conductivity, $\kappa_T$ is proportional to density, and also rather weakly dependent on temperature (this distinction will be of fundamental importance in the sequel). Although it is distinctly possible that the turbulent heat conductivity depends on quantities such as the magnetic energy release rate (via the fluctuation energy $\delta B_{\perp}^{2}/8\pi$) or on the magnetic correlation length $\lambda_{B}(t)$, we here, for simplicity, take $\lambda_{T}$ to be a constant parameter.  Based on analysis of hard X-ray observations, in particular the ratio of intensities of coronal and chromospheric hard X-ray sources \citep{2013A&A...551A.135S}, \cite{2014ApJ...780..176K} have suggested that $\lambda_T \simeq 10^8$~cm.  For temperatures $T \simeq 10^7$~K and densities $n \simeq 10^{10}$~cm$^{-3}$, the collisional mean free path $\lambda_{ei} \simeq 10^8$~cm (Equation~(\ref{lambdaei})) and so values $\lambda_B \gapprox 10^8$cm do not result in meaningful differences from earlier results \citep[e.g.][]{1995ApJ...439.1034C} based on collision-dominated conduction. In order to adequately explore the effects of turbulent heat flux limitation, we therefore here model $\lambda_T$ in the range $(10^6$ -- $5 \times 10^7)$~cm.  We shall find that a value $\lambda_T \simeq 10^7$~cm best accounts for the observed \citep{2013ApJ...778...68R} cooling times.

The scattering frequency $\nu$ for an electron of velocity $v$ is $v/\lambda$, where $\lambda$ is the mean free path associated with the particular scattering process under consideration. In the presence of both collisional and turbulent scattering, the scattering frequencies are generally additive, and therefore the mean free path $\lambda$ is given by

\begin{equation}\label{lambda_general}
\frac{1}{\lambda} = \frac{1}{\lambda_{ei}} + \frac{1}{\lambda_T} \,\,\, .
\end{equation}
It is useful to define the turbulent reduction factor

\begin{equation}\label{R-T}
R(T, n ; \lambda_{T}) = \frac{\lambda_{ei}}{\lambda_T} = \frac{\kappa_C}{\kappa_T}= \frac{(2k_B T)^2}{2\pi e^4 \ln \Lambda \, n} \, \frac{1}{\lambda_T} \equiv \eta \, \frac{T^{2}}{n\lambda_{T}} \,\,\, ,
\end{equation}
where $\eta \equiv 2 k_B^2/\pi e^4 \ln \Lambda = 1.14 \times 10^4$~cm$^{-2}$~K$^{-2}$.  In the collision-dominated regime $\lambda_{ei} \ll \lambda_T$ ($R \ll 1$), the heat flux takes the usual form

\begin{equation}\label{fc-collisional}
F_C = - \kappa_0 \, T^{5/2} \, \frac{dT}{dz} = - \frac{2}{7} \, \kappa_{0} \, \frac{dT^{7/2}}{dz} \simeq - \frac{2}{7} \, \kappa_{0} \, \frac{T_a^{7/2}}{L} \,\,\, ,
\end{equation}
where $T_a$ is the apex temperature and $L$ is the loop half length. On the other hand, in the turbulence-dominated regime, $\lambda_{T}\ll \lambda_{ei}$ ($R \gg 1$), and the heat flux is given by

\begin{equation}\label{fc-turbulent}
F_T = - \frac{\kappa_0 \, T^{5/2}}{R(T, n; \lambda_T)} \, \frac{dT}{dz} = - \frac{n \, \lambda_T \kappa_0 \, T^{1/2}}{\eta} \, \frac{dT}{dz} = - \frac{2}{3} \, \frac{n\lambda_{T}\kappa_{0}}{\eta} \, \frac{dT^{3/2}}{dz} \simeq - \frac{2}{3} \, \frac{\kappa_0}{R(T_a) } \, \frac{T_a^{7/2}}{L} \,\,\, .
\end{equation}
Consideration of the limiting forms~(\ref{fc-collisional}) and~(\ref{fc-turbulent}) suggests the following approximate analytic form for the conductive heat flux, valid in both the low-$R$ (high $\lambda_T$) and high-$R$ (low $\lambda_T$) regimes:

\begin{equation}\label{conduction-approximate}
F_c = - \frac{2} {7 + 3 \, R(T_a, n; \lambda_{T})} \,\, \kappa_0 \, \frac{T_a^{7/2}}{L} \,\,\, .
\end{equation}
This is the form of the heat flux that we shall use in solving the EBTEL equations that describe the evolution of temperature and density in the cooling post-flare loop.

\subsection{Development of the EBTEL equations}

The main idea behind the zero-dimensional (0-D), loop-integrated, EBTEL formalism \citep{2008ApJ...682.1351K,2012ApJ...752..161C,2012ApJ...758....5C} is that the enthalpy flux associated with mass exchange between the coronal and chromospheric parts of the loop is at all times in balance with the excess (or deficit) of the heat flux relative to the transition region radiation loss rate. When the conductive heat flux exceeds transition region radiative losses, the excess heat flux is deposited in the chromosphere, leading to a substantial increase in gas pressure; the resulting pressure-driven ``evaporation'' of chromospheric plasma increases the density of the coronal and transition regions of the loop until the enhanced radiative losses ($\propto n^2$) can now balance the downward heat flux from the corona. On the other hand, when the heat flux is in deficit, the loop plasma cools radiatively and drains back into the chromosphere, reducing the density to a level where the radiative loss rate is now in balance with the thermal conductive flux.  The loop-averaged density and temperature thus depend on a balance between thermal conduction and radiation, a balance that one expects to shift when the heat flux is suppressed by turbulent angular-scattering processes.  Evaluating the extent of this shift is the main purpose of the present work.

As shown by \cite{2008ApJ...682.1351K}, the EBTEL enthalpy balance formalism leads to the following equations  describing the time evolutions of the average pressure and temperature along a given magnetic field line (``strand''):

\begin{eqnarray}\label{ebtel-p-n}
\frac{1}{\gamma-1} \, \frac{dp}{dt} & = & - \frac{R_C}{L} \, (1+c_1) + Q(t) \cr \cr
\frac{dn}{dt} & = & - \frac{c_2}{c_3} \, \frac{(\gamma-1)}{2 \gamma \, k_{B} T L } \, (F_c + c_1 \, R_C) \,\,\, ,
\end{eqnarray}
where $p=2nk_BT$ (erg~cm$^{-3}$) is the pressure, $n$ (cm$^{-3}$) is the density, and $T$ (K) is the electron temperature (all averaged over the extent of the loop), $\gamma$ is the ratio of specific heats (here taken to be 5/3), and $L$ is the loop half-length. $Q(t)$ (erg~cm$^{-3}$~s$^{-1}$) is the volumetric heating rate and

\begin{equation}\label{radiation-expression}
R_C = n^2 \, \Lambda(T) \, L
\end{equation}
(erg~cm$^{-2}$~s$^{-1}$) is the (optically thin) radiative loss per unit cross-sectional area, i.e., the integral of the volumetric radiative energy losses over a loop half-length.  Here we take $\Lambda(T)$ (erg~cm$^3$~s$^{-1}$) in the form $\Lambda (T) = \zeta T^{\alpha} = 1.2\times10^{-19} \, T^{-1/2}$, which, in the pertinent range of temperatures ($10^6$~K -- $5 \times 10^7$~K), gives values very similar to the $\Lambda(T) = 1.95 \times 10^{-18} \, T^{-2/3}$ relation used by \cite{2008ApJ...682.1351K}.

The constants $c_1$, $c_2$ and $c_3$ in Equations~(\ref{ebtel-p-n}) are here taken as in the revised EBTEL paper \citep{2012ApJ...752..161C}. $c_{1}$ is the ratio of radiative losses from the transition region of the loop to those from the coronal region and varies between 2 and 0.6 throughout the simulation (see Equation~\eqref{c1-variation} below), $c_2 = 0.89$ is the ratio of the average coronal temperature to the maximum temperature at the loop apex, and $c_3 = 0.6$ is the ratio of the base coronal temperature to the apex temperature.  Hence, according to Equation~(\ref{conduction-approximate}), the conductive flux is represented in terms of the average loop temperature $T$ by

\begin{equation}\label{fc-ebtel}
F_c = - \frac{2}{7 + 3 \, R(T/c_2, n; \lambda_T)} \, \frac{\kappa_0}{L} \, \left ( \frac{T}{c_2} \right )^{7/2} \,\,\, .
\end{equation}

We can eliminate $dn/dt$ between Equations~(\ref{ebtel-p-n}) to obtain an expression for the rate of change of temperature

\begin{equation}\label{ebtel-t}
\frac{dT}{dt} = \frac{1}{n k_B} \left \{ - \, \frac{2 c_2}{5 \, c_3 \, L^2} \,\, \frac{\kappa_0}{7 + 3 \, R(T/c_{2},n,\lambda_{T})} \, \left(\frac{T}{c_2}\right)^{7/2} - \left [ \frac{1+c_{1}}{3} - \frac{c_1 \, c_2}{5 \, c_3} \right ] \, n^{2} \, \zeta \, T^{-1/2} + \frac{1}{3} \, Q(t) \right \} \,\,\, ;
\end{equation}
this, plus the second of Equations~(\ref{ebtel-p-n}), viz.

\begin{equation}\label{ebtel-n}
\frac{dn}{dt} = \frac{1}{k_B T} \, \frac{c_2}{5 \, c_3} \, \left \{ \frac{2}{L^2} \,\, \frac{\kappa_0}{7 + 3 \, R(T/c_{2},n,\lambda_{T})} \left(\frac{T}{c_2}\right)^{7/2} - c_1 \, n^2 \, \zeta \, T^{-1/2} \right \} \,\,\, ,
\end{equation}
govern the evolution of the loop-averaged temperature and density, respectively.

It is convenient to introduce the auxiliary constants

\begin{equation}\label{a-collisional}
a = \frac{2 \kappa_0}{35 \, k_{B} \, c_3 \, c_{2}^{5/2} \, L^2} \,\,\, ,
\end{equation}

\begin{equation}\label{b-collisional}
b = \left [  \frac{1+c_1}{3} \, - \frac{ c_1 \, c_2}{5 \, c_3} \right ] \, \frac{\zeta}{k_B} \,\,\, ,
\end{equation}
and

\begin{equation}\label{d-collisional}
d = \frac{ c_{1} \, c_{2}}{5 \, c_3} \, \frac{\zeta}{k_B} \,\,\, ,
\end{equation}
which allow Equations~(\ref{ebtel-t}) and~(\ref{ebtel-n}) to be rewritten in the more readable forms:

\begin{equation}\label{ebtel-t-abd}
\frac{dT}{dt} = - \frac{a}{1+ \frac{3}{7} \, R(T/c_2, n; \lambda_T)} \, \frac{T^{7/2}}{n} - b \, n \, T^{-1/2} + \frac{Q(t)}{3 \, k_B \, n} \,\,\, ,
\end{equation}
and

\begin{equation}\label{ebtel-n-abd}
\frac{dn}{dt} = \frac{a}{1+ \frac{3}{7} \, R(T/c_2, n; \lambda_{T})} \, T^{5/2} - d \, n^{2}{T^{-3/2}} \,\,\, .
\end{equation}

Equations~(\ref{ebtel-t-abd}) and~(\ref{ebtel-n-abd}) form the basis for the subsequent analysis of this paper.  They extend the hydrodynamically static equations considered by \citet{2016ApJ...833...76B} to include the effect of mass motions in addition to {\it in situ} plasma heating/cooling.  The numerical values and units of the various constants and parameters entering these equations, plus the assumed value ($10^9$~cm $\simeq 15\arcsec$) for the loop half-length $L$ and initial values for temperature ($T_0$) and density ($n_0$), are displayed in Table~\ref{table:1}; note that the values of $b$ and $d$ depend on the value of $c_1$, which varies between an initial value of 2 to a value as low as 0.6 throughout the simulation (see Equation~\eqref{c1-variation}) below).

\begin{table}[H]
\begin{center}
\begin{tabular}{||c c||}
\hline
Constant & Value \\ [0.5ex]
\hline\hline
$c_1$ & 0.6 - 2.0\\
\hline
$c_2$ & 0.89 \\
\hline
$c_3$ & 0.6 \\
\hline
$\kappa_{0}$ & $1.7\times10^{-6}$~erg~cm$^{-1}$~s$^{-1}$~K$^{-7/2}$ \\
\hline
$\zeta$ & $1.2\times10^{-19}$~erg cm$^3$~s$^{-1}$~K$^{1/2}$ \\
\hline
$a$ & $1.57 \times 10^{-9}$~cm$^{-3}$~s$^{-1}$~K$^{-5/2}$ \\
\hline
$b$ & $(3.10 - 3.54) \times 10^{-4}$~cm$^{3}$~s$^{-1}$~K$^{3/2}$ \\
\hline
$d$& $(1.55 - 5.16) \times 10^{-4}$~cm$^{3}$~s$^{-1}$~K$^{3/2}$ \\
\hline \hline
Parameter & Value \\ [0.5ex]
\hline\hline
$L$ & $1 \times 10^9$~cm \\
\hline
$T_0$ & $1.5 \times 10^7$~K \\
\hline
$n_0$ & $1 \times 10^{10}$~cm$^{-3}$ \\
\hline
\end{tabular}
\caption{Values of Constants and Parameters}
\label{table:1}
\end{center}
\end{table}

\section{Conductive and radiative cooling regimes}\label{cooling-regimes-overview}

We first consider Equations~(\ref{ebtel-t-abd}) and~(\ref{ebtel-n-abd}) in the absence of heating (i.e., $Q(t)=0$), and solve them with initial conditions ($T=T_0, n=n_0$ at $t=t_0$) in limiting regimes.

\subsection{Conductive cooling regime}

If conditions are such that thermal conduction dominates the energy balance, then we may neglect the radiative terms, casting Equations~(\ref{ebtel-t-abd}) and~(\ref{ebtel-n-abd}) into the form

\begin{equation}\label{dtdt-conduction}
\frac{dT}{dt} = -\frac{a}{1+ \frac{3}{7} \, R(T/c_{2},n;\lambda_{T})} \,\, \frac{T^{7/2}}{n} \,\,\, ,
\end{equation}
and

\begin{equation}\label{dndt-conduction}
\frac{dn}{dt} = \frac{a}{1+ \frac{3}{7} \, R(T/c_{2},n;\lambda_{T})} \,\,  T^{5/2} \,\,\, .
\end{equation}
(Note that the constants $a$ does not depend on $c_1$, so that the results in the conduction-dominated regime do not depend on the value of $c_1$ chosen.)  Multiplying the first of these by $n$, the second by $T$, and adding yields $d(nT)/dt=0$, so that the constant pressure condition

\begin{equation}\label{pres}
n(t) \, T(t) = n_0 T_0
\end{equation}
holds.  This result follows from the fact that thermal conduction merely redistributes energy throughout the loop volume, so that the average energy density $3 n k_B T$ (and also the pressure $p=2 n k_B T$) does not change with time. (It also follows trivially from the first of Equations~(\ref{ebtel-p-n}) when both $R_C$ and $Q(t)$ are set equal to zero.)  Taking the reciprocal of Equation~(\ref{dtdt-conduction}), and using Equations~(\ref{R-T}) and~(\ref{pres}), yields an (implicit) analytic solution for $T(t)$ (and hence $n(t)$) for all values of $\lambda_T$. Here we consider only the solutions in the limit of large $\lambda_T$ (collision-dominated conduction) and small $\lambda_T$ (turbulence-dominated conduction).

\subsubsection{Collision-dominated}

For large values of $\lambda_{T}$ (small values of $R$), corresponding to collision-dominated conduction, Equations~(\ref{dtdt-conduction}) and~(\ref{dndt-conduction}) reduce, using Equation~(\ref{pres}), to

\begin{equation}
\frac{dT}{dt} = - \, \frac{a}{n_0 T_{0}} \, T^{9/2} \, ; \qquad \frac{dn}{dt} = a \, (n_0 T_0)^{5/2} \, n^{-5/2} \,\,\, .
\end{equation}
These have solutions

\begin{equation}\label{t-n-solution-conduction-only}
T(t) = T_{0} \, \left [ 1 + \frac{(t-t_0)}{\tau_{cC}} \right ]^{-2/7} \, ; \qquad n(t)= n_{0} \, \left [ 1 +\frac{(t-t_0)}{\tau_{cC}} \right ]^{2/7} \,\,\, ,
\end{equation}
where the collision-dominated conduction time

\begin{equation}\label{tau-cs}
\tau_{cC} = \frac{2 n_0}{7 a T_0^{5/2}} \equiv
\frac{5 \, c_3 \, c_2^{5/2} n_0 k_B L^2}{\kappa_0 T_0^{5/2}} \,\,\, .
\end{equation}
For comparison, Equations~(26) and~(27) of \citet{2016ApJ...833...76B} give the evolution of temperature and density for collision-dominated conductive cooling in a {\it static} loop:

\begin{equation}\label{temperature-evolution-static-collisional}
T(t) = T_{0} \, \left [ 1 + \frac{(t-t_0)}{\tau_{cC,S}} \right ]^{-2/5} \, ; \qquad n = n_0  \, \,\, ,
\end{equation}
where the static-model collision-dominated conduction time

\begin{equation}\label{tau-css}
\tau_{cC,S} = \frac{21 n_0 k_B L^2}{20 \, \kappa_0 T_0^{5/2}} = \frac{21}{100 \, c_3 \, c_2^{5/2}} \, \tau_{cC} \simeq \frac{\tau_{cC}}{2} \,\,\, .
\end{equation}
The smaller exponent in the expression for $T(t)$ in Equation~(\ref{t-n-solution-conduction-only}) compared to that in the static solution~(\ref{temperature-evolution-static-collisional}), coupled with the factor of two larger characteristic cooling time (Equation~(\ref{tau-css})), show that the temperature evolves more slowly with time than it does in the static case.  Indeed, the initial cooling in the EBTEL model can be approximated by

\begin{equation}\label{t-initial-cooling}
T(t) \simeq T_0 \left [ 1 - \frac{2}{7} \, \frac{(t-t_0)}{\tau_{cC}} \right ] \,\,\, ; \qquad n(t) \simeq n_0 \left [ 1 + \frac{2}{7} \, \frac{(t-t_0)}{\tau_{cC}} \right ] \,\,\, ,
\end{equation}
while in the static case

\begin{equation}\label{t-initial-cooling-static}
T(t) \simeq T_0 \left [ 1 - \frac{2}{5} \, \frac{(t-t_0)}{\tau_{cC,S}} \right ] \simeq T_0 \left [ 1 - \frac{4}{5} \, \frac{(t-t_0)}{\tau_{cC}} \right ] \,\,\, ; \qquad n(t) = n_0 \,\,\, .
\end{equation}
This significantly more gradual decay of temperature in the EBTEL scenario compared to the static scenario is due to the flow of enthalpy into a cooling volume in order to maintain the constant pressure condition.

\subsubsection{Turbulence-dominated}

On the other hand, for small $\lambda_T$, corresponding to turbulence-dominated conduction, Equations~(\ref{dtdt-conduction}) and~(\ref{dndt-conduction}) reduce to

\begin{equation}
\frac{dT}{dt} = - \frac{7 c_2^2 \lambda_T}{3 \eta} \, a \, T^{3/2} \, ; \qquad \frac{dn}{dt} = \frac{7 c_2^2 \lambda_T}{3 \eta} \, (n_0 T_0)^{1/2} \, a \, n^{1/2} \,\,\, .
\end{equation}
These have solutions

\begin{equation}\label{t-n-solution-turbulent-conduction}
T(t)= T_0 \left [ 1 +  \frac{(t-t_0)}{\tau_{cT}} \right ]^{-2} \, ; \qquad n (t)= n_0 \left [ 1 +  \frac{(t-t_0)}{\tau_{cT}} \right ]^2 \,\,\, ,
\end{equation}
where the turbulent conductive time

\begin{equation}\label{tauct}
\tau_{cT} = \frac{6 \eta}{7 \, a \, c_{2}^2 \, \lambda_T T_0^{1/2}} \equiv \frac{15}{2^{3/2}} \, c_3 \, c_2^{1/2} \, \left ( \frac{m_e}{k_B T_0} \right )^{1/2} \, \frac{L^2}{\lambda_T} \,\,\, .
\end{equation}

Equations~(\ref{t-n-solution-conduction-only}) in the collision-dominated case, or Equations~(\ref{t-n-solution-turbulent-conduction}) in the turbulence-dominated case, both show that during the conductive phase of cooling the density increases such that the pressure (thermal energy) remains approximately constant. This is the phenomenon of ``gentle evaporation'' first described by \citet{1978ApJ...220.1137A}.

For comparison, Equations~(32) and~(33) of \citet{2016ApJ...833...76B} give the following variations for a static cooling model in the turbulence-dominated conduction regime:

\begin{equation}\label{temperature-evolution-static-turbulent}
T(t)= T_0 \left [ 1 +  \frac{(t-t_0)}{\tau_{cT,S}} \right ]^{-2} \, ; \qquad n = n_0 \,\,\, ,
\end{equation}
where the static-model turbulence-dominated conduction time

\begin{equation}\label{tau-cts}
\tau_{cT,S} = \frac{9}{2^{7/2}} \, c_3 \, c_2^{1/2} \, \left ( \frac{m_e}{k_B T_0} \right )^{1/2} \, \frac{L^2}{\lambda_T} = \frac{3}{20 \, c_3 \, c_2^{1/2}} \, \tau_{cT} \simeq \frac{\tau_{cT}}{4} \,\,\, .
\end{equation}
Once again, the cooling rate in the EBTEL model is substantially slower than in the static model.  Although the exponents in Equations~(\ref{t-n-solution-turbulent-conduction}) and~(\ref{temperature-evolution-static-turbulent}) are the same, the characteristic cooling time for the EBTEL case is about four times larger than in the case of collision-dominated conduction.

\subsection{Radiative cooling regime}

We now consider the complementary case where the conductive terms are negligible compared to the radiative terms in Equations~(\ref{ebtel-t-abd}) and~(\ref{ebtel-n-abd}), which then reduce to

\begin{equation}\label{ebtel-t-collisional-no-conduction}
\frac{dT}{dt} = - b \, \frac{n}{T^{1/2}} \,\,\, ,
\end{equation}
and

\begin{equation}\label{ebtel-n-collisional-no-conduction}
\frac{dn}{dt} = - d \, \frac{n^2}{T^{3/2}} \,\,\, .
\end{equation}
A radiation-dominated scenario typically becomes pertinent at later stages of the loop cooling process. Therefore we express the solutions of Equations~(\ref{ebtel-t-collisional-no-conduction}) and~(\ref{ebtel-n-collisional-no-conduction}) with initial conditions $T = T_{*}, n = n_{*}$ at time $t =t_{*}$ corresponding to the temperature and density at the transition from conduction-dominated cooling to radiative-dominated cooling. From Equations~(\ref{ebtel-t-collisional-no-conduction}) and~(\ref{ebtel-n-collisional-no-conduction}),

\begin{equation}
\frac{1}{n} \, \frac{dn}{dt} = \frac{d}{b} \, \frac{1}{T} \, \frac{dT}{dt} \,\,\, ,
\end{equation}
so that

\begin{equation}\label{nTgb-constant}
n \propto T^{d/b} \,\,\, ,
\end{equation}
with (Equations~(\ref{b-collisional}) and~(\ref{d-collisional}))

\begin{equation}\label{d-over-b}
\frac{d}{b} \equiv \frac{1}{ \frac{5 \, c_3}{3 \, c_2} \, \left ( 1+\frac{1}{c_1} \right ) -1 } \,\,\, .
\end{equation}
The value of $d/b$ depends on the value chosen for the constant $c_1$.  As argued by \cite{2012ApJ...752..161C} (their Section~3), for a near-uniform pressure loop $c_1$ starts at a value $\simeq 2$ (their Equation~(11)) and then transitions to a lower value in the radiative phase.  Using the observed \citep{1991A&A...241..197S} scaling $T \propto n^2$ (i.e., $d/b = 1/2$), we find from Equations~(\ref{nTgb-constant}) and~(\ref{d-over-b}) that $c_1 = 0.6$ is the appropriate value for the radiative phase.  We therefore use the expression

\begin{equation}\label{c1-variation}
c_1 = \frac{2 + 0.6 \left ( \frac{n}{n_{bal}} \right )^2}{1 + \left ( \frac{n}{n_{eq}} \right )^2 } \,\,\,  ,
\end{equation}
where $n_{eq}$ is the density required for density equilibrium ($dn/dt = 0$) at temperature $T$.  From Equation~(\ref{ebtel-n}) (and using Equation~(\ref{R-T})),

\begin{equation}\label{n-bal}
n_{eq} = \frac{T^2}{2c_2^2 \lambda_T} \left  [ \sqrt{ \left ( \frac{3 \eta}{7} \right )^2 + 4 \, \frac{a}{d} \, c_2^4 \, \lambda_T^2} - \frac{3 \eta}{7} \right ]
\end{equation}
(compare with Equation~(17) of \citep{2012ApJ...758....5C} for the case of Spitzer conductivity ($\lambda_T \rightarrow \infty$)).  Equation~(\ref{c1-variation}) is comparable to Equation~(18) of \cite{2012ApJ...752..161C}, but has been modified slightly to produce the limiting values $c_1=2$ for low values of $n$ (conduction-dominated phase) and $c_1=0.6$ for the high values of $n$ pertinent to the radiation-dominated phase \citep{1991A&A...241..197S}.

Using Equation~(\ref{nTgb-constant}), Equation~(\ref{ebtel-t-collisional-no-conduction}) may be written as

\begin{equation}\label{T-radiative-t-only}
\frac{dT}{dt} = -b \, \frac{n_{*T}}{T_{*T}^{d/b}} \, T^{\frac{d}{b} - \frac{1}{2}} \,\,\, .
\end{equation}
The form of the solution to this equation depends crucially on the value of $d/b$. When $d/b<3/2$ the solution is

\begin{equation}\label{t-db-lt-32}
T=T_{*} \left [ 1-\frac{(t-t_{*})}{\tau_{r}} \right ]^{\frac{1}{3/2-d/b}}  ; \qquad
\tau_{r}=\frac{T_{*}^{3/2}}{n_{*}b \left ( \frac{3}{2} - \frac{d}{b} \right )} \,\,\, .
\end{equation}
We can in this case give a clear interpretation to the cooling time $\tau_{r}$: it is the time (beyond $t=t_*$) that it takes for the temperature to reach zero. On the other hand, for $d/b>3/2$ the solution is

\begin{equation}\label{t-db-gt-32}
T=T_{*} \left [ 1+\frac{(t-t_{*})}{\tau_{r}} \right ]^{\frac{1}{3/2-d/b}} ; \qquad \tau_{r}=\frac{T_{*}^{3/2}}{n_{*}b \left ( \frac{d}{b} - \frac{3}{2} \right )} \,\,\, ,
\end{equation}
so that the temperature evolution bifurcates from a concave-down to a concave-up function of time. When $d/b = 3/2$ exactly, the temperature evolution is exponential ($dT/dt = T/\tau_r$, with $\tau_r = T_*^{3/2}/n_* b$).  This exponential (Newton) cooling profile can be obtained easily from either of Equations~(\ref{t-db-lt-32}) or~(\ref{t-db-gt-32}) using the
relation

\begin{equation}
e^{-t/\tau_r}=\lim_{\kappa \rightarrow \infty} \left ( 1 \pm \frac{t}{\kappa \, \tau_r} \right )^{\mp \kappa},
\end{equation}

Interestingly, for the particular case studied here ($d/b=1/2$),
the radiative cooling profile is {\it linear}:

\begin{equation}\label{t-radiative-solution}
T(t) = T_{*} \left [ 1 -  \frac{(t - t_{*})}{\tau_{r}} \right ]\, ; \qquad \tau_{r} \simeq \frac{3.3 \times 10^3 \, T_{*}^{3/2}}{n_{*}} \,\,\, .
\end{equation}
We thus see that the phenomenological thermodynamical scaling law $T \propto n^{2}$ (corresponding to $d/b = 1/2$), originally proposed by \cite{1991A&A...241..197S} and subsequently found by other authors \citep{1995ApJ...439.1034C,2005A&A...437..311B,2010ApJ...710L..39B,2010ApJ...717..163B} to apply to short coronal loops, corresponds to a linear cooling profile. This point, already made by
\cite{1995ApJ...439.1034C}, is therefore also a feature of the EBTEL model. However, it has been proposed \citep{2010ApJ...717..163B} that for long, tenuous loops, the correct scaling is instead $T\propto n$, corresponding to $d/b=1$.  Equation~(\ref{t-db-lt-32}) shows that such a scaling law is associated with a cooling profile which has a concave down parabolic behavior $T = T_0 \, [1 - (t-t_*)/\tau_r]^2$, reaching zero temperature in a finite time $t = t_* + \tau_r$.  It can also be noted that for the value $c_1=2$ used in the original EBTEL work \citep{2008ApJ...682.1351K}, $d/b \simeq 5$, which results in a concave-up temperature profile $T = T_0 \, [1 + (t-t_*)/\tau_r]^{-2/7}$ in which the temperature approaches zero asymptotically.

For the density evolution, Equation~(\ref{ebtel-n-collisional-no-conduction}) may be written as

\begin{equation}
\frac{dn}{dt} = - d \, \frac{n_{*}^{3b/2d}}{T_{*}^{3/2}} \, n^{2-3b/2d} \,\,\, .
\end{equation}
For $d/b<3/2$, the solution is

\begin{equation}\label{n-radiative-solution-dblt32}
n(t) = n_{*} \left [ 1 -  \frac{(t - t_{*})}{\tau_{r}} \right ]^{\frac{d/b}{3/2-d/b}} \, ; \qquad \tau_{r} = \frac{T_{*}^{3/2}}{n_{*}b \left ( \frac{3}{2} - \frac{d}{b} \right )} \,\,\, ,
\end{equation}
so that again the exponent is positive and the function is concave down. The solutions for $d/b>3/2$ are similarly
obtained by a simple change of sign, so that the exponent becomes negative and the function becomes concave up.

For the pertinent case $d/b=1/2$, we have

\begin{equation}\label{n-radiative-solution}
n(t) = n_{*} \left [ 1 -  \frac{(t - t_{*})}{\tau_{r}} \right ]^{1/2} \, ; \qquad \tau_{r} \simeq \frac{3.3\times 10^{3} \, T_{*}^{3/2}}{n_{*}} \,\,\, .
\end{equation}
In the radiative cooling phase the density thus decreases with decreasing temperature: $n \simeq T^{1/2}$, corresponding to radiative draining of the loop and the \cite{1991A&A...241..197S} scaling. The pressure decreases according to

\begin{equation}\label{T-n-5}
p(t) = p_{0} \left [ 1 -  \frac{(t - t_{*T})}{\tau_{r}} \right ]^{3/2} \ .
\end{equation}

Equation~(\ref{t-radiative-solution}) can be compared with Equation~(39) of \cite{2016ApJ...833...76B} for the static solution.  Setting the exponent of temperature in the radiative loss rate $\ell = 1/2$, their result is

\begin{equation}\label{t-radiative-static}
T(t) = T_{*T} \left ( 1 - \frac{(t-t_{*T})}{\tau_r} \right )^{2/3} \, ; \qquad \tau_r \simeq 2.5 \times 10^3 \, \frac{T_{*T}^{3/2}}{n_{*T}} \,\,\, ,
\end{equation}
with a very similar cooling time and profile (and a slightly concave down behavior).

\section{Cooling scenarios}\label{cooling-scenarios}

\begin{figure}[H]
\begin{center}
\includegraphics[width=0.9\textwidth]{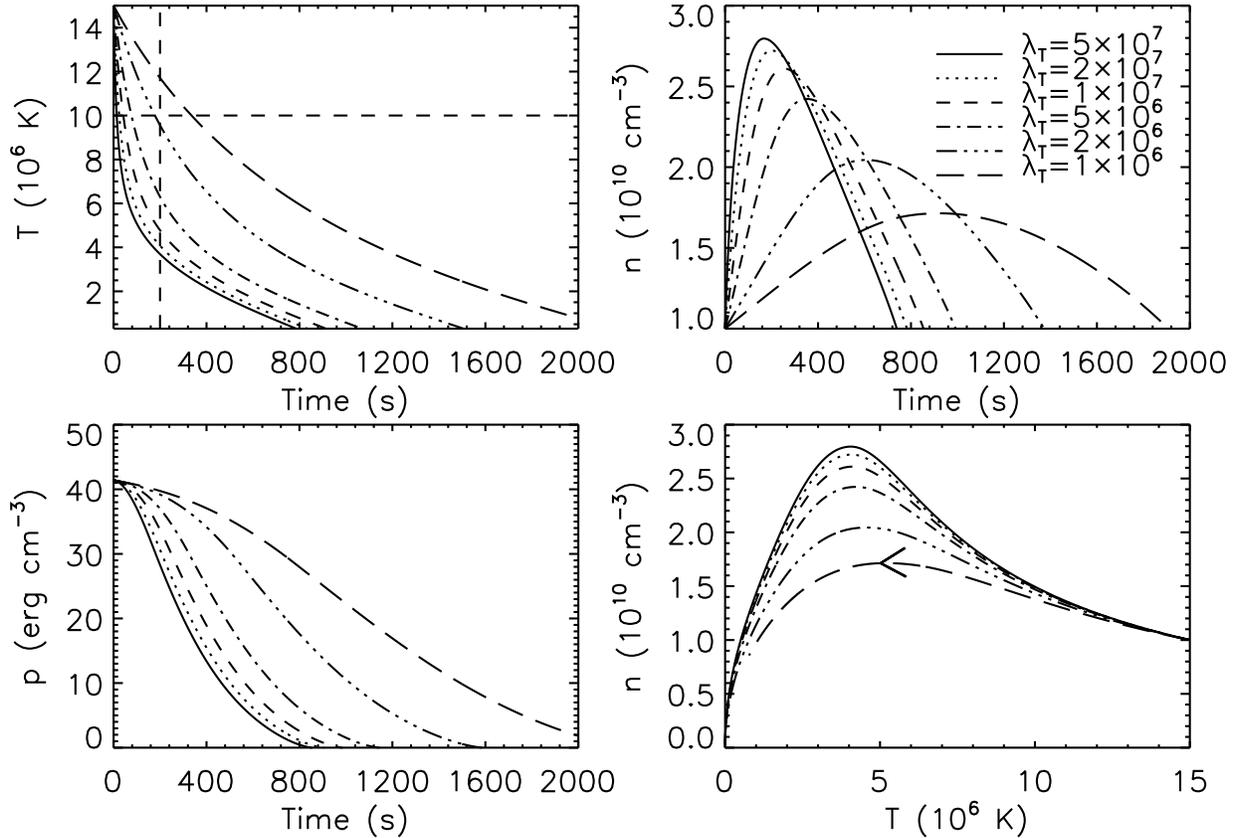}
\caption{Variation of temperature (top left panel), density (top right panel), and pressure (bottom left panel), for cooling models associated with various values of the turbulent mean free path $\lambda_T$.  A $(T,n)$ phase plot of the various solutions appears in the bottom right panel.  The dashed lines in the top right panel show that for $\lambda_T = 2 \times 10^6$~cm, the temperature does indeed cool to $10^7$~K in about 200~s, consistent with the observations of \cite{2013ApJ...778...68R}.}\label{cooling-profiles}
\end{center}
\end{figure}

We now explore numerically the temperature and density evolution of the loop from an initial temperature $T_0 = 1.5 \times 10^7$~K and density $n_0 = 1 \times 10^{10}$~cm$^{-3}$ (Table~\ref{table:1}). The temperature, density and pressure profiles, together with a phase plot ($T,n$) of the solution, are shown in Figure~\ref{cooling-profiles}, for various values of the turbulence mean free path $\lambda_T$.

For large $\lambda_{T}$ values, Coulomb collisions dominate the thermal conduction term during the entire cooling process. Starting from a temperature $T_0$ and density $n_0$, the loop starts to cool down, initially by thermal conduction of heat toward the chromosphere.  The pressure gradients established by this chromospheric heating drive an upward (evaporative) motion from the chromosphere toward the corona leading to a rise in density in order to maintain an approximately constant pressure (Equation~(\ref{pres})) in the loop. In due course, the temperature becomes small enough and the density high enough such that the cooling becomes dominated by radiation.  Thereafter the temperature continues to decrease (but on a time scale less than the conductive time scale) and the loop steadily drains the evaporated material back to the chromosphere.  However, the cooling time from $T_0 = 1.5 \times 10^7$~K to $T = 1 \times 10^7$~K is very short; indeed, for the threshold value $\lambda_T = \eta T_0^2/c_2^2 n_0 \simeq 7 \times 10^8$~cm that produces $R=1$ (Equation~(\ref{R-T})) and hence allows turbulence to start playing a role in the cooling process, the cooling time is only $\simeq5$~s, which is much shorter than the $\sim 200$~s obtained from observations \citep{2013ApJ...778...68R}.

Figure~\ref{cooling-profiles} shows that a value $\lambda_T \simeq 2 \times 10^6$~cm corresponds to a cooling time from $T_0 = 1.5 \times 10^7$~K to $T = 1 \times 10^7$~K that is consistent with the $\simeq200$~s observed time \citep{2013ApJ...778...68R}.  For comparison, \citet{2016ApJ...833...76B} found that $\lambda_T \simeq 5 \times 10^6$~cm gave the best fit to the observations in a static cooling model; the (factor of 3) smaller value of $\lambda_T$ found here is because of the hydrodynamic terms in the EBTEL model. Physically, because the initial increase in density in a hydrodynamic model leads to a greater radiative cooling term, a greater turbulent reduction factor $R$, and hence a smaller value of $\lambda_T$, is required to contain more heat in the corona and so produce the same temperature-versus-time profile.  The inferred value of $\lambda_T$ is, however, significantly smaller than the value $\lambda_T \simeq 10^8$~cm estimated by \cite{2014ApJ...780..176K} from considerations of the variation of hard X-ray source size with energy.

\section{Influence of the heating term}\label{heating-function}

We now consider the role and effect of the heating terms in the EBTEL equations

\begin{equation}\label{laterT}
\frac{dT}{dt} = -\frac{a}{1+ \frac{3}{7} R(T/c_{2},n;\lambda_{T})} \, \frac{T^{7/2}}{n} - b \, nT^{-1/2} + \frac{Q_B+Q(t)}{3k_{B}n} \,\,\, ,
\end{equation}

\begin{equation}\label{latern}
\frac{dn}{dt} = \frac{a}{1+ \frac{3}{7} R(T/c_{2},n;\lambda_{T})} \, T^{5/2} - d \, n^{2}{T^{-3/2}} \,\,\, .
\end{equation}
Here $Q_B$ is the background heating necessary to maintain the quiescent loop at temperature $T_0$, and $Q(t)$ is a flare heating term, which we shall see below must play an important role, even in the loop cooling phase.

\subsection{Preflare steady state}\label{steady-state}

Pre-flare, the loop is in a steady state with a temperature $T_{eq,0}$ of, say, $10^6$~K and, since it precedes any impulsive release of energy, also corresponds to a regime of very limited (if any) turbulence.  We therefore take the $R=0$ limit of Equation~(\ref{latern}) to obtain

\begin{equation}\label{neq-no-turbulence}
n_{eq,0}=\sqrt{\frac{a}{d}} \,\, T_{eq,0}^2 \simeq 1.8 \times 10^9 \, {\rm cm}^{-3} \,\,\, ,
\end{equation}
a value that is consistent with actual preflare (active region) conditions. Then, from the energy equation~(\ref{laterT}), also in the low-$R$ limit, we have

\begin{equation}\label{Qb-full}
\frac{dT}{dt} = 0 \Rightarrow Q_{B} = 3k_B \left ( a \,  T_{eq,0}^{7/2} + b \, \frac{n_{eq,0}^2}{T_{eq,0}^{1/2}} \right ) = 3 k_B a \, \left ( 1 + \frac{b}{d} \right ) \, T_{eq,0}^{7/2} \simeq 10^{-3} \, {\rm erg \, cm}^{-3} \, {\rm s}^{-1}  \,\,\, ,
\end{equation}
so that only a tiny fraction ($\simeq 10^{-6}$) of the available magnetic energy density is required to be dissipated each second.

\subsection{Flare heating}

The flare heating kernel $Q(t)$ (erg~cm$^{-3}$~s$^{-1}$) is taken to have a Gaussian time profile

\begin{equation}\label{heating-func}
Q(t) = \frac{q}{\Delta t \sqrt{\pi}} \,\,  e^{-\frac{(t-t_q)^2}{\Delta t^2}} \,\,\, ,
\end{equation}
where

\begin{equation}
q=\int Q(t) \, dt
\end{equation}
is the total magnetic energy density released per unit volume over the duration of the heating, $\Delta t$ is the duration of energy release, and $t_{q}$ the time of peak heating. The strength $q$ and duration $\Delta t$ of the heating must be such that the following constraints are satisfied:

\begin{enumerate}

\item at some time after the start of the energy release, the temperature and density should reach $T=T_{0} \simeq 1.5 \times 10^7$~K and $n=n_{0} \simeq 10^{10}$~cm$^{-3}$ ;
\item the loop must cool from a state with $T \simeq 1.5 \times 10^7$~K and $n \geq 10^{10}$~cm$^{-3}$ down to $10^7$~K in $\simeq200$~s in order to be consistent with observations \citep[e.g.,][]{2013ApJ...778...68R};
\item the energy released per unit volume $q$ must be less than the available magnetic energy density.  For definiteness, we consider a magnetic field $B_0 = 300$~G, so that, assuming \citep[see, e.g.,][]{2004JGRA..10910104E} that some 30\% of the magnetic energy is excess over the ground-state potential field and so available for conversion,

\begin{equation}\label{available-magnetic}
q \approx 0.3 \, \frac{B_0^2}{8\pi} \simeq 10^3~{\rm erg~cm}^{-3} \,\,\, .
\end{equation}
Over a volume $\sim L^3 \simeq 10^{27}$~cm$^{3}$, this corresponds to a total released energy of $10^{30}$~erg, consistent with the decay-phase heating amounts quoted in the last column of Table~1 in \cite{2013ApJ...778...68R}.

\end{enumerate}

We have explored various different values of $q$ and $\Delta t$ in the Gaussian heating function $Q(t)$, and different values of the turbulent mean free-path $\lambda_{T}$, in an attempt to satisfy all these constraints. We have also explored different functional forms for $Q(t)$, and found that the essential results depend not so much on the exact shape of $Q(t)$, but rather on its characteristic amplitude $q$ and width $\Delta t$.  We below describe the essential features of these models using some illustrative models, labeled in terms of the values of $q$, $\Delta t$, and $\lambda_T$:

\begin{equation}\label{model-definition}
M(q;\Delta t;\lambda_{T}).
\end{equation}

\subsubsection{Instantaneous heat pulse}
In the limit $\Delta t \rightarrow 0$,

\begin{equation}\label{delta-pulse}
Q(t) \rightarrow q \, \delta(t-t_q) \,\,\, ,
\end{equation}
where it should be noted that $t_{q}$, the time of peak heating, does not have to coincide with the target time $t_{0}$ at which $T(t)=T_{0}\simeq 1.5 \times 10^{7}$~K and $n(t)=n_{0}\simeq 10^{10}$~cm$^{-3}$. In this extreme case of impulsive heating, the temperature increases instantaneously to a value $T_{q}$ and, since evaporative mass motions have no time to develop, the density $n_{q}$ remains equal to its preflare equilibrium value:

\begin{equation}
n_{q}=n_{eq,0}.
\end{equation}
Of course, such a heating model cannot satisfy constraint $\#1$ unless $t_q < t_0$. Further, the value of $q$ which will raise the temperature from $T_{eq,0}$ to $T_q$ is straightforwardly computed:

\begin{equation}
q=3 k_B n_{eq,0} \, (T_q - T_{eq,0}) \,\,\, .
\end{equation}
In absence of turbulent limitation of heat conduction in the preflare atmosphere, $n_{eq,0} \simeq 2 \times 10^{9}$~cm$^{-3}$ (Equation~(\ref{neq-no-turbulence})), and hence $q\simeq 12~{\rm erg~cm}^{-3}$ is required to raise the temperature from $T_{eq}\simeq 10^6$~K to $T_{q}\simeq 1.5\times 10^7$~K. This is a tiny fraction $\simeq 1\%$ of the total available (non-potential) magnetic energy. Now we know from Section~\ref{cooling-regimes-overview} that during the conductive cooling phase the pressure remains (approximately) constant (Equation~(\ref{pres})), and hence that

\begin{equation}
n_{0}T_{0} = n_{q}T_{q} = n_{eq,0} T_q \,\,\, .
\end{equation}
Thus, in order to satisfy constraint $\#2$, i.e., to obtain a temperature $T_0 \simeq 1 \times 10^7$~K and a density $n_0 \simeq 10^{10}$~cm$^{3}$ at some later time $t_0$ during the cooling phase, then $T_q/T_0 = n_0/n_{eq,0} \simeq 5$. We thus see that while a $\delta$-impulse heating model can in principle satisfy all three constraints above, they will generally be associated with a rather high value of the peak temperature $T_q \simeq 8 \times 10^7$~K, as we now illustrate numerically.

\begin{figure}[H]
\begin{center}
\includegraphics[width=0.9\textwidth]{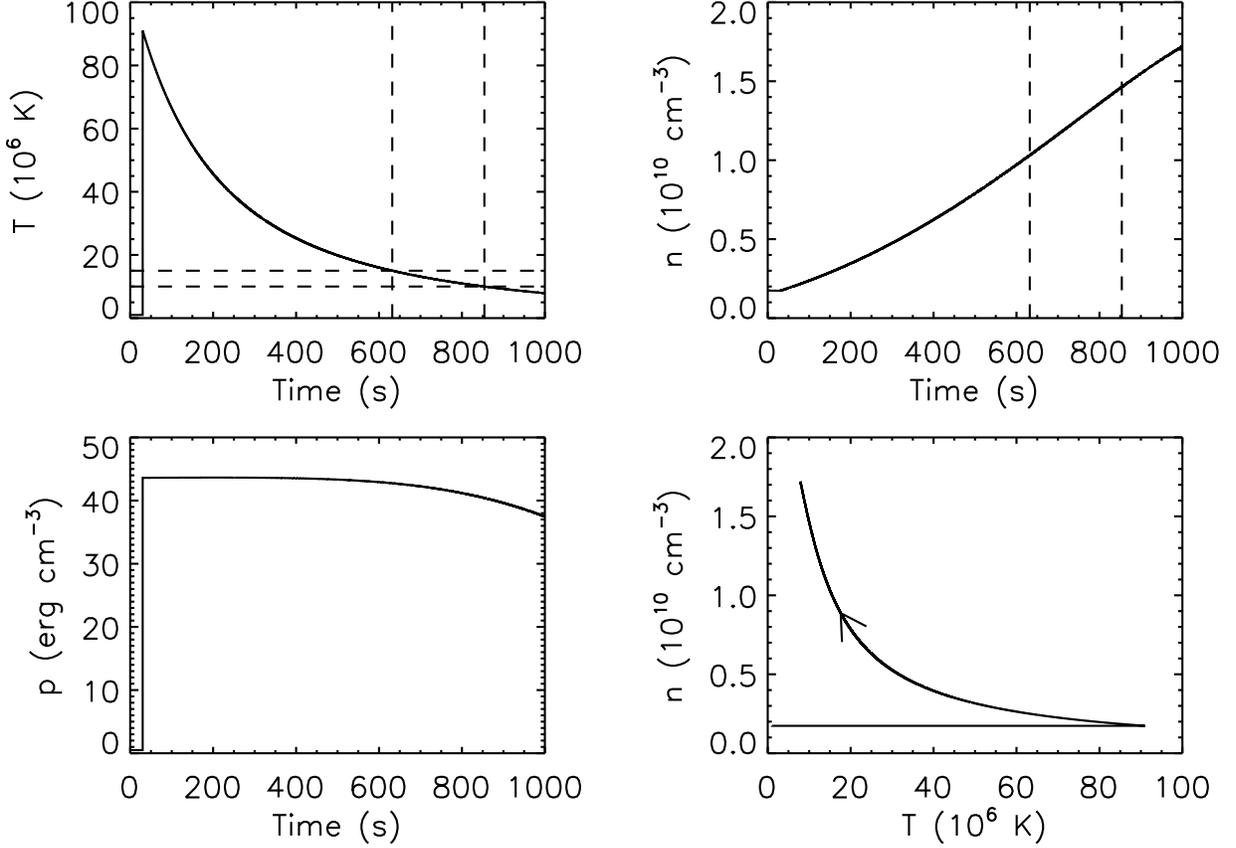}
\caption{{\it M($65$; $0$; $2 \times 10^{6}$)}: Response of a turbulent loop to a $\delta$-function heat pulse with $q = 65$~erg~cm$^{-3}$, $t_0 = 30$~s, and $\lambda_T = 2 \times 10^6$~cm; the panels are essentially the same as in Figure~\ref{cooling-profiles}.  The peak temperature, produced at the same time as the energy injection, is quite large ($T_{q}\simeq 2 \times 10^8$~K). As shown by the dashed lines in the top two panels, about 600~s after the time of energy input the loop has cooled to $\simeq 1.5 \times 10^7$~K, with its density rising to $\simeq 10^{10}$~cm$^{-3}$. Further cooling to $\simeq 10^{7}$~K takes $\simeq 220$~s, consistent with the observations of \cite{2013ApJ...778...68R}, and during this cooling period the density increases to $\simeq 1.5 \times 10^{10}$~cm$^{-3}$.  The pressure (lower left panel) remains roughly constant throughout the first 1000~s of the simulation. The lower right panel shows an $(n,T)$ phase plot of the solution.}\label{Model-3-fig}
\end{center}
\end{figure}

\underline {Model $M(65; 0; 2 \times 10^6$)}.
We carried out a series of simulations using $\lambda_T = 2 \times 10^{6}$~cm and found (Figure~\ref{Model-3-fig}) that a $\delta$-function heat pulse with $q\simeq 65$~erg~cm$^{-3}$ did simultaneously produce a temperature $T = 1.5 \times10^7$~K and a density $n \simeq 10^{10}$~cm$^{-3}$, in addition to reproducing the correct cooling time (Constraint \#2). However, as expected above, this model produces a very large peak temperature $T_q \simeq 9 \times 10^7$~K. Further, the required ``initial'' temperature $T_0$ and density $n_0$ are reached only after quite a long time $t \simeq 600$~s after the time of energy input. The pressure remains roughly constant throughout the first 1000~s or so, due to the relatively low radiative losses corresponding to such low densities.

\subsubsection{Extended Gaussian heat pulse}

The failure of impulsive heating models with the correct cooling time to account for the observed peak temperature shows that the heating function $Q(t)$ must extend over a finite time {\it and therefore will play a significant role in the behavior of temperature during the cooling phase}.  We now consider extended Gaussian heating functions $Q(t)$ (Equation~(\ref{heating-func})) to see if a scenario consistent with all three constraints can be obtained. We can already anticipate that increasing $\Delta t$ for a prescribed $q$ will result in a decrease of the peak temperature. Therefore, in order to achieve sufficiently large peak temperatures, and hence densities (constraint $\#1$), as well as reproducing the correct cooling time (constraint $\#2$), it will be necessary to deposit a much larger energy density $q$ in the loop.  Accordingly, as we now illustrate, such extended heating models will therefore become constrained by $\#3$, namely that the energy release $q$ must not exceed the total magnetic energy density available.

\begin{figure}[H]
\begin{center}
\includegraphics[width=0.9\textwidth]{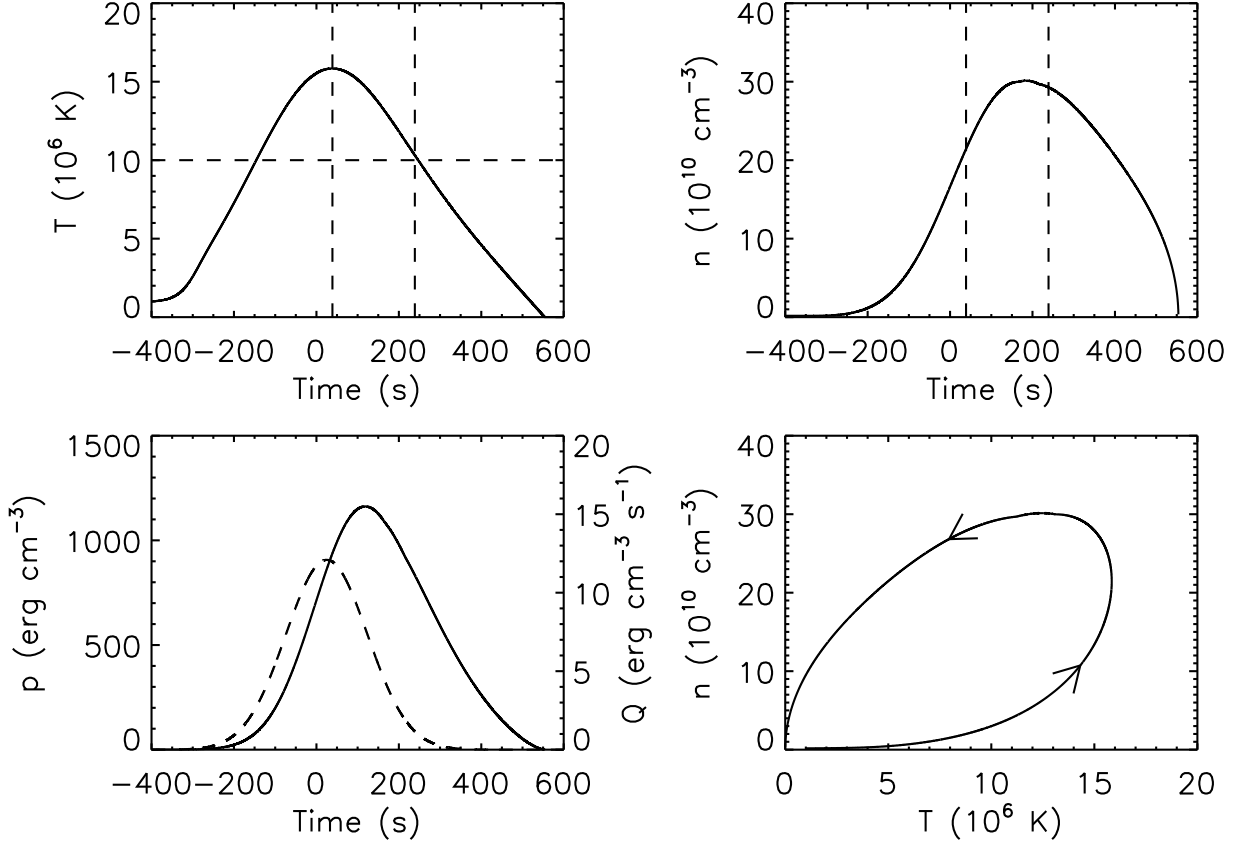}
\caption{{\it M($3000$,$140$, $\infty$)}: Response of the loop to a Gaussian heat pulse with $q \simeq 3 \times 10^3$~erg~cm$^{-3}$ and $\Delta t \simeq 140$~s (dashed curve in lower left panel), with $\lambda_T = \infty$ (no turbulent limitation of heat conduction). The panels are the same as in Figure~\ref{Model-3-fig}. The temperature rises to $T \simeq 1.6 \times 10^7$~K a few seconds after the peak energy release, and the loop then cools to $1 \times 10^7$~K in $\simeq 200$~s (dashed lines in top left panel), consistent with the observations of \cite{2013ApJ...778...68R}.  However, the relatively large amount of heat deposited causes a substantial amount of chromospheric heating and hence a large amount of chromospheric evaporation.  Accordingly, the density continues to rise to a large value $\simeq 3 \times 10^{11}$~cm$^{-3}$. The lower left panel shows the evolution of the pressure, which (Equation~(\ref{ebtel-p-n})) is $(\gamma-1)$ times the energy density and evolves approximately as the time integral of the heat input $Q(t)$.  The energy density peaks at a very high value $\simeq 1750$~erg~cm$^{-3}$, greater than the total available magnetic energy density (Equation~(\ref{available-magnetic})).}\label{Model-5-fig}
\end{center}
\end{figure}

\underline {Model $M(3000, 140, \infty$)}. Figure~\ref{Model-5-fig} shows that taking $q = 3 \times 10^3$~erg~cm$^{-3}$ and $\Delta t=140$~s, with $\lambda_T = \infty$ (no turbulent limitation of heat conduction) yields a peak temperature of $T = 1.6 \times 10^7$~K.  Further, the cooling time to $10^{7}$~K is (top left panel of Figure~\ref{Model-5-fig}) is $\simeq 200$~s, consistent with observations \citep{2013ApJ...778...68R}. Over the same period the relatively large amount of heat deposited exceeds the ability of the transition zone plasma to radiate it away.  The excess heat is deposited in the chromosphere where it creates a substantial amount of chromospheric evaporation, causing the density to continue to rise from $\simeq 2 \times 10^{11}$~cm$^{-3}$ to $\simeq 3 \times 10^{11}$~cm$^{-3}$.  As expected, the pressure ($=(\gamma-1) \, \times$ the energy density) behaves roughly as $(\gamma -1) \times$ the integral of the energy input $Q(t)$ (dashed line in lower left panel of Figure~\ref{Model-5-fig}), peaking near the end of the energy input profile.  The peak energy density reaches a value $\simeq 1750$~erg~cm$^{-3}$, comparable to the total energy deposited (the remainder of the deposited energy is radiated away). Such a model is fully consistent with observations; however, the required value of $q \simeq 3 \times 10^3$~erg~cm$^{-3}$ now corresponds to a release of more than the entire available magnetic energy density in the loop (Equation~(\ref{available-magnetic})), in violation of Constraint \#3.

\subsubsection{Extended Gaussian heat pulse with heat flux limitation}

The above results show further that it is very challenging for a extended heating model that involves collision-dominated heat conduction to produce sufficiently large values of the loop temperature and density (Constraint \#1) and a cooling time consistent with observations (Constraint \#2), while releasing an acceptably small amount of the available magnetic energy (Constraint \#3).  {\it To meet all the constraints listed above, we must include the turbulent suppression of heat conduction}; as we now illustrate.

\begin{figure}[H]
\begin{center}
\includegraphics[width=0.9\textwidth]{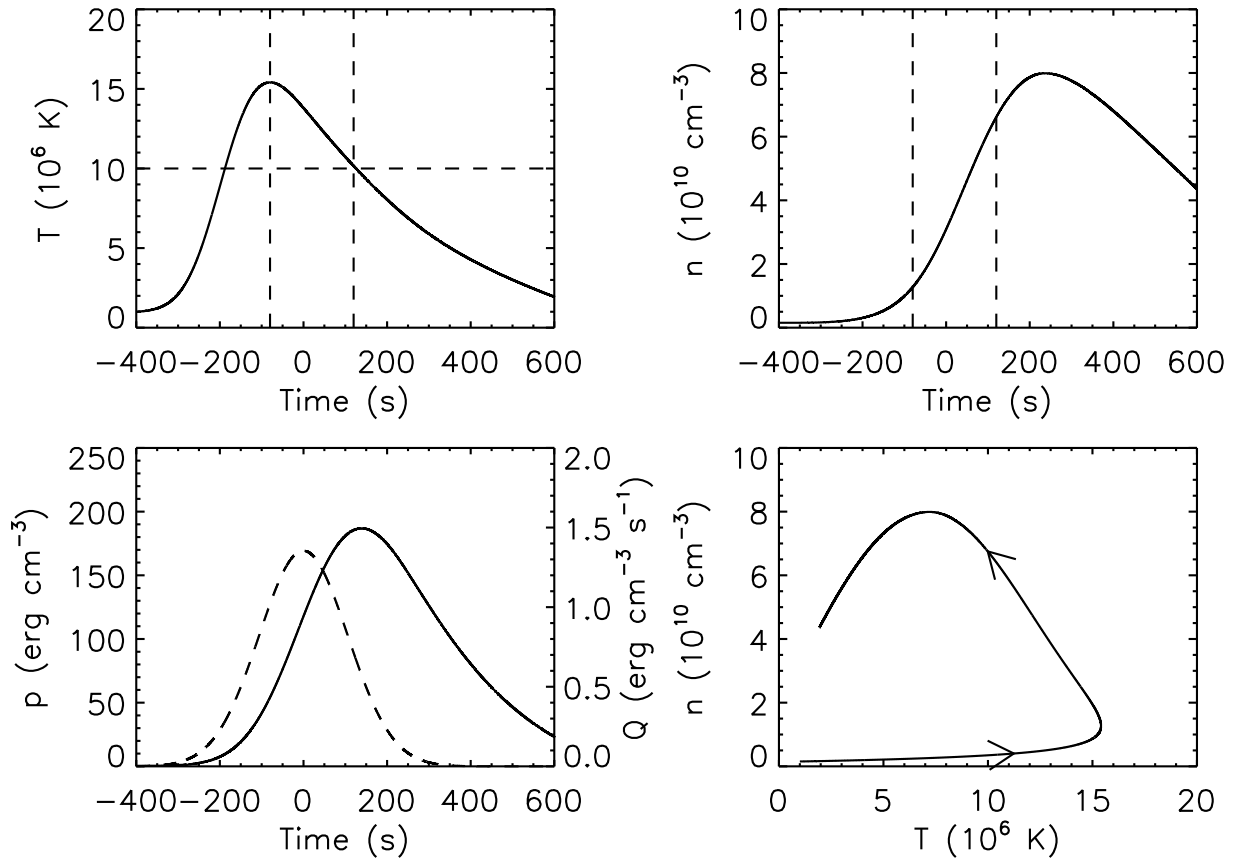}
\caption{{\it $M(360, 150, 1.4 \times 10^7$)}: Response of the loop to a Gaussian heat pulse with $q = 360$~erg~cm$^{-3}$, $\Delta t = 150$~s, and $\lambda_T = 1.4 \times 10^7$~cm.  The panels are the same as in Figure~\ref{Model-3-fig}. The dashed lines in the top left panel show the cooling from $1.5 \times 10^7$~K to $10^7$~K in 200~s, consistent with the observations of \cite{2013ApJ...778...68R}.  During the same 200~s period, the density rises from $\simeq 10^{10}$~cm$^{-3}$ to $\simeq 8 \times 10^{10}$~cm$^{-3}$ (upper right panel). The lower left panel shows the energy input profile $Q(t)$ (dashed line) and the loop pressure $p$ (solid line); again the energy density $(=3p/2)$ behaves approximately as the time integral of the energy input $Q(t)$.  The peak pressure of $\simeq 190$~erg~cm$^{-3}$~s$^{-1}$ corresponds to an energy density $\simeq 280$~erg~cm$^{-3}$~s$^{-1}$, or about one-fourth of the available magnetic energy (Equation~(\ref{available-magnetic})). Such a model therefore satisfies all of the imposed constraints.}\label{Model-6-fig}
\end{center}
\end{figure}

\underline {Model $M(360, 150, 1.4 \times 10^7$)}. Taking a heat pulse centered on $t_{q}=0$~s, with $q \simeq 360$~erg~cm$^{-3}$ ($\simeq 30\%$ of the total non-potential magnetic energy density available; Constraint \#3), $\Delta t = 150$~s, and $\lambda_T = 1.4 \times 10^7$~cm, we obtain a peak temperature $T \simeq 1.5 \times 10^7$~K that is reached $80$~s {\it before} the peak of the energy input, with a corresponding density $n \simeq 1.3 \times 10^{10}$~cm$^{-3}$.  The time taken to further cool to $T = 1 \times 10^7$~K is $\simeq 200$~s (see dashed vertical and horizontal lines in top left panel of Figure~\ref{Model-6-fig}), consistent with Constraint \#2.  During this time the loop density rises steadily to $n \simeq 8 \times 10^{10}$~cm$^{-3}$. The peak energy density ($=3p_{max}/2$) $\simeq 280$~erg~cm$^{-3}$, or about one quarter of the available (non-potential) magnetic energy (Equation~(\ref{available-magnetic})), consistent with Constraint \#3. This model thus satisfies all of the imposed constraints. The required value $\lambda_T \simeq 1.4 \times 10^7$~cm is about an order of magnitude larger than the value $\lambda_T \simeq 2 \times 10^6$~cm required to produce the correct cooling profile in the absence of continued energy release (Section~\ref{cooling-scenarios}), and is closer to the value $\lambda_T \simeq 10^8$~cm estimated by \cite{2014ApJ...780..176K} from hard X-ray observations.

\section{Summary and conclusions}\label{conclusions}

We have used the ``enthalpy-based thermal evolution of loops'' (EBTEL) model to study the hydrodynamics of a flaring loop in which heat conduction is limited by turbulent scattering of thermal electrons, using a theory originally developed by \cite{2016ApJ...824...78B}. The motivation for such a study lies in the long-standing observations of anomalously large cooling times in the
post-impulsive phase of solar flares \citep[e.g., ][]{1980sfsl.work..341M} and recent observations carried by \cite{2017PhRvL.118o5101K} which suggest that the magnetic energy released into acceleration of non-thermal electrons is associated with a surge of turbulence.  If the turbulence generated by the magnetic reconnection energy release persists beyond the impulsive phase, it can confine the conductive heat flux and hence result in much longer cooling times consistent with those observed \citep{1980sfsl.work..341M,2013ApJ...778...68R}

In the spirit of the work by \cite{1995ApJ...439.1034C}, we have solved the EBTEL equations piece-wise analytically in both the conduction-dominated and radiation-dominated regimes of cooling.  The resulting expressions show that in order to account for the observed cooling times, classical collisional transport must play at best a limited role in cooling the high-temperature plasma created by a solar flare.  We have also shown through numerical simulation (Section~\ref{heating-function}) that heat-flux-confining turbulence cannot by itself account for the observed temperatures, emission measures and cooling times; an extended period of magnetic energy release is also required.  Furthermore, however, invoking an extended period of energy input cannot by itself produce consistency with observations while simultaneously satisfying the rather evident constraint that the total energy released over time must be a relatively small fraction of the available non-potential magnetic energy.  Combined, these results show that {\it extended duration energy release} and {\it the presence of turbulence} are two essential facets related to energy release and transport in solar flares.

Finally, we have shown that a model characterized by a turbulent mean free path $\lambda_T \simeq 10^7$~cm does a remarkably good job in explaining the peak temperatures and densities in flaring loops and the observed lengthy cooling times, while requiring only $\simeq$30\% of the available magnetic energy to be released.  This inferred value of $\lambda_T$ is comparable to that inferred from considerations of electron transport in connection with hard X-ray observations \citep{2014ApJ...780..176K}.

\acknowledgments

We thank the referee for several very valuable comments, both qualitative and quantitative.  NB and AGE were supported by Grant NNX17AI16G from the NASA Heliophysics Supporting Research program. The work of EPK and NB was also partially supported by STFC consolidated grant ST/L000741/1.

\bibliographystyle{apj}
\bibliography{ms}

\end{document}